\documentclass[reqno]{article}
\usepackage{amssymb,amsmath,cite,delarray,epsfig}
\begin{document}

\begin{flushright}
{{OSU-HEP-02-11}}\\
{{UMD-PP-03--002}}\\
{{SLAC-PUB-9274}}
\end{flushright}
\begin{center}
~
\vskip.6in
{\LARGE \bf Radiative Processes ($\tau \rightarrow \mu\gamma$,
$\mu \rightarrow e\gamma$ and $(g-2)_\mu$)}\\[0.1in]
{\LARGE\bf as Probes of ESSM/SO(10)}
\vskip1.0in
{\large\bf K. S. Babu}$^{(a)}$ and {\large\bf Jogesh C. Pati}$^{(b),(c)}$
\\[0.2in] 
$^{(a)}${Department of Physics,  Oklahoma State University,
Stillwater, OK 74078, USA}\\
$^{(b)}${Department of Physics, 
University of Maryland, College Park,  MD 20742, USA\footnote{Permanent
address}}\\
$^{(c)}${Stanford Linear Accelerator Center, 
Menlo Park, CA 94025, USA}
\end{center}

\vskip1.0in
\begin{abstract}

The Extended Supersymmetric Standard Model (ESSM), motivated on several
grounds, introduces two vectorlike families (${\bf 16}+{\bf \bar{16}}$)
of $SO(10)$) with masses of order one TeV.  It is noted that the
successful predictions of prior work on fermion masses and mixings,
based on MSSM embedded in $SO(10)$, can be retained rather simply
within the ESSM extension.  These include an understanding of the
smallness of $V_{cb} \approx 0.04$ and the largeness of $\nu_\mu-\nu_\tau$
oscillation angle, $\sin^22\theta_{\nu_\mu \nu_\tau}^{\rm osc} \approx 1$.  
We analyze the new contributions arising through the exchange 
of the vectorlike families of ESSM to radiative processes including
$\tau \rightarrow \mu \gamma$, $\mu \rightarrow e \gamma$, $b \rightarrow
s \gamma$, EDM of the muon and the muon $(g-2)$.  We show that ESSM
makes significant contributions especially to the decays $\tau \rightarrow
\mu \gamma$ and $\mu \rightarrow e \gamma$ and simultaneously to muon
$(g-2)$.  For a large and plausible range of relevant parameters, we
obtain: $a_\mu^{ESSM} \approx +(10-40) \times 10^{-10}$, with
a correlated  prediction that $\tau \rightarrow \mu \gamma$ should be
discovered with an improvement in its current limit by a factor of
3-20.  The implications for $\mu \rightarrow e\gamma$ are very
similar.  The muon EDM is within reach of the next generation
experiments.  Thus, ESSM with heavy leptons being lighter than about
700 GeV (say) can be probed effectively by radiative processes
before a direct search for these vectorlike leptons and quarks
is feasible at the LHC.

\end{abstract}
\newpage
{\large
\section{Introduction}
A variant of MSSM -- the so-called Extended Supersymmetric Standard Model
(ESSM) -- has been motivated  sometime ago on several grounds
\cite{JCPBabu,BabuJCP}. Briefly speaking, in addition to the three chiral
families, ESSM introduces two vectorlike families of quarks and leptons
(together with their superpartners) that transform as
{\bf 16}+${\bf \bar{16}}$ of SO(10), and possess an SO(10)-invariant mass of
order one TeV. It assumes that the three chiral families
acquire their masses primarily (barring small corrections of order one MeV)
through their mixings with the two vectorlike families. As we will explain, this
mechanism of mass-generation for the three chiral families has the advantage
that it provides a simple understanding of the interfamily mass-hierarchy
($m_{u,d,e}\ll m_{c,s,\mu}\ll m_{t,b,\tau}$) \cite{JCPBabu,BabuJCP}. In
particular, it automatically renders the electron family massless (barring
small corrections $\sim$ 1 MeV) and also naturally accounts for the
$\mu/\tau$ mass-hierarchy, even if no small numbers are introduced from the
start.

In the sequel we will list other theoretical motivations for the ESSM tied
to issues that arise in the context of unification, and also the reason for
its consistency with LEP neutrino counting as well as precision electroweak
tests. No doubt the vectorlike quarks and leptons, if they exist with masses
$\lesssim$ 1-2 TeV, as ESSM proposes, would be visible prominently at the LHC.
Recently it has been noted \cite{BP2002} that ESSM with the heavy lepton
members having masses $\lesssim$ 500 GeV (say), would provide a simple
explanation of the anomaly in $\nu N$-scattering that has been recently
reported by the NuTeV group \cite{NuTeV} and simultaneously of the LEP
neutrino counting that is presently at 2$\sigma$ below the standard model
value of 3 \cite{LEPcount}.

The purpose of this note is to stress that radiative processes -- in
particular $\tau \rightarrow \mu\gamma$, $\mu \rightarrow e\gamma$ and muon
$(g-2)$ associated with the vertex $\mu \rightarrow \mu\gamma$ -- can provide
additional sensitive probes of ESSM. Of these three processes, the
measurement of muon $(g-2)$ has drawn special attention over the last year.
This is because, the BNL result based on its 1999-data \cite{BNL}, in spite
of the realization of the reversal of sign of the hadronic light-by-light
scattering contribution to $(g-2)_\mu$ \cite{light}, points to a
possible anomaly in $(g-2)_\mu$, given by $\delta a_\mu\equiv a_\mu^{\rm expt}-
a_\mu^{SM}\approx (25\pm 16)\times 10^{-10}$. This result by itself would
suggest that $\delta a_\mu$ could quite possibly lie in the range of
(10-40)$\times 10^{-10}$. Such a view has recently been called to question,
however, in Ref. \cite{MusolfWise}, where it is noted that the hadronic
light-by-light scattering contribution to $(g-2)_\mu$ has a rather large
uncertainty given by $(\pm 6 +3\tilde c)\times 10^{-10}$. While model
calculations yield $\tilde c\approx 1$ \cite{Model}, in general $\tilde c$
is expected to be of order unity with either sign. In the presence of such
uncertainty, a definitive conclusion as to whether there exists an anomaly
[$a_\mu\gtrsim \times 10^{-10}$ (say)] would have to await a further
reduction of experimental error (which is due soon), as well as (depending
upon the central value) a reduction in the theoretical uncertainty of
hadronic effects. Meanwhile, anticipating \underline{either} outcome, it
seems worthwhile to explore possible new physics which would contribute to
$a_\mu$ in the range mentioned above, especially if such physics is motivated
on other grounds. Theoretical exploration of this kind could eventually help
constrain new physics regardless of whether the final verdict confirms or
denies an anomaly in $(g-2)_\mu$.

It has been noted by several authors \cite{SUSY} that low energy-supersymmetry
\cite{FN11new} arising in the context of MSSM is a natural source of the new
contribution to $a_\mu$. As we will show in this paper, ESSM would provide
an additional source of new contribution to $a_\mu$, which can naturally be
in the range of (10-40)$\times 10^{-10}$, provided the heavy leptons are
relatively light ($m_{E,E'}\approx M_{N,N'}\approx 250$-500 GeV, say).
The intriguing feature of ESSM with such moderately light heavy leptons is
that it leads to crucial predictions as regards observability of especially
$\tau\rightarrow\mu\gamma$ and also $\mu\rightarrow e\gamma$. In this sense,
ESSM with a moderately light spectrum would be testable even
before LHC turns on.

We recall some salient features of ESSM and theoretical motivations in its
favor in the next section. In Sec. \ref{sect2} we discuss the Yukawa
couplings and fermion mass matrices for the case of ESSM to indicate how one
can essentially reproduce in this case the successes of the
G(224)/SO(10)-framework for fermion masses and mixings that was presented in
Ref. \cite{BPW}, for the case of MSSM. In Sec. \ref{sect3} we use this
realistic framework to discuss the contributions of ESSM to $(g-2)_\mu$,
$\tau\rightarrow\mu\gamma$, $\mu\rightarrow e\gamma$, the muon
electric dipole moment and $b\rightarrow s \gamma$. In Sec. 5
we present a summary and concluding remarks.

\section{Salient Features of ESSM}
\label{sect1.5}
The so called "Extended
Supersymmetric Standard Model" (ESSM), which introduces {\it two complete
"vectorlike" families of quarks and leptons} -- denoted by
$Q_{L,R}=(U,D,N,E)_{L,R}$ and $Q'_{L,R}=(U',D',N',E')_{L,R}$ -- with
relatively light masses of order one TeV. Both
$Q_L$ {\it and} $Q_R$ transform as (2,1,4), while $Q'_L$ and $Q'_R$
transform as (1,2,4) of the symmetry group
G(224)=SU(2)$_L\times$SU(2)$_R\times$SU(4)$^C$. Thus together they have
the quantum numbers of a pair of
{\bf 16}+${\bf \bar{16}}$ of SO(10), to be denoted by
${\bf 16_V}=(Q_L|\bar{Q}'_R)$ and
${\bf \bar{16}_V}=(\bar{Q}_R|Q'_L)$. The subscript "V" signifies two
features:
(a) ${\bf \bar{16}_V}$ combines primarily with ${\bf 16_V}$, so that the
pair gets a
(dominant) SO(10)-invariant mass term of the form
$M_V{\bf 16_V\cdot\bar{16}_V}+
h.c.=M_V(\bar{Q}_RQ_L+\bar{Q}'_RQ'_L)+h.c.$, at the GUT scale, presumably
utilizing the VEV of an SO(10)-singlet (see below). (b) Since $Q_L$ and
$Q_R$ are doublets of SU(2)$_L$, the massive four-component object
$(Q_L\oplus Q_R)$ couples vectorially to $W_L$'s; likewise
$(Q'_L\oplus Q'_R)$ couples vectorially to $W_R$'s. Hence the name
"$\underline{\mbox{vectorlike}}$" families. The three chiral families are
denoted by $(16_i)$, $i=1,2,3$.

It is assumed (see e.g., Ref. \cite{JCPBabu} and \cite{BabuJCP}) that
the mass term $M_V$ of the two vectorlike families is protected by some
local generalized "flavor" or discrete symmetries (presumably of string
origin), so that it is of order TeV, rather than the GUT-scale, just like
the $\mu$-term of MSSM. It is furthermore assumed that the same set of
"flavor" symmetries dictate that the direct mass term of the three chiral
families which could arise from couplings of the form $h_{ij}^{(0)}
{\bf 16_i16_j}\Sigma_H$ (where $\Sigma_H={\bf 10_H}$ or
${\bf 10_H\times 45_{H}}/M$, etc.), are
strongly suppressed, up to small corrections $\lesssim$ a few MeV
(see remarks later).
Thus the
chiral families get their masses (barring corrections $\lesssim$ a few MeV)
primarily through their mixings with the two vectorlike families. It is shown
in the next section that such a pattern of the $5\times 5$
mass matrix, involving the three chiral and the two vectorlike families,
would naturally yield an exactly massless family (barring corrections
$\lesssim$ a few MeV) and an inter-family mass-hierarchy ($m_{u,d,e}\ll
m_{c,s,\mu}\ll m_{t,b,\tau}$), even if such large hierarchy ratios were not
present to begin with in the mass-elements that mix the three chiral with the
two vectorlike families \cite{JCPBabu,BabuJCP,Barr90}.

A few general comments about ESSM are in order. Note that it of course
preserves all the merits of MSSM as regards gauge coupling unification
and protection of the Higgs masses against large quantum corrections. From
the point of view of adding extra families of quarks and leptons, ESSM in
fact seems to be the minimal as well as the maximal extension of the MSSM
that
is allowed by (a) LEP neutrino counting, (b) precision measurements
of the oblique electroweak parameters \cite{PeskinAltarelli, BPZhang} as well
as (c) the demand of a perturbative or semi-perturbative
\cite{BabuJCP, HemplingKoldaMarr} as opposed to a nonperturbative
gauge coupling unification [e.g., addition of a fourth
chiral family, as opposed to two vectorlike families of ESSM, would in
general be incompatible with (b)].

Theoretical motivations for the case of ESSM arise on several grounds: (a)
It raises $\alpha_{\rm unif}$ to a semiperturbative value of 0.25 to 0.3 and
therefore provides a much better chance to stabilize the dilaton than
the case of MSSM, for which $\alpha_{\rm unif}$ is rather weak (only 0.04)
\cite{BabuJCP}; (b) Owing to increased two-loop effects, ESSM raises the
unification scale $M_X$ to about (0.5-2)$\times 10^{17}$ GeV
\cite{BabuJCP,HemplingKoldaMarr} and thereby considerably reduces the
problem of a mismatch between the MSSM and the string unification scales
\cite{Ginspang, RevDenesJCP}; (c) It lowers the GUT-prediction for
$\alpha_3(m_Z)$ compared to that for MSSM \cite{BabuJCP}, as needed by the
data \cite{PleData,Langacker}; (d) It naturally enhances the GUT-prediction
for proton lifetime \cite{BPWJCP} compared to that for MSSM embedded in a
GUT, also as needed by the data (i.e., by the SuperK limit); and finally
 (e) as noted above, ESSM provides a simple reason for inter-family mass
 hierarchy \cite{JCPBabu,BabuJCP,Barr90}.

 In this sense, ESSM, though less economical than MSSM, offers some distinct
 advantages over MSSM. The main purpose of this paper is to point
 out that ESSM can also offer a simple explanation of the muon
 $(g-2)$ anomaly, should it eventually persist, without requiring a light
 SUSY threshold. Simultaneously,
 it would offer a set of crucial tests, involving radiative processes,
 especially $\tau\rightarrow\mu\gamma$, and also $\mu\rightarrow e\gamma$,
 and edm of the muon and
 last but not least a clear potential for the discovery of a host of
 vectorlike quarks and leptons, in addition to the SUSY particles, at the
 LHC and possibly the NLC.

 In the next section we discuss the Yukawa coupling and fermion mass matrices
 for the case of ESSM to indicate how one can essentially reproduce in this
 case the successful SO(10)-framework for fermion masses and mixings that was
 presented in Ref. \cite{BPW} for the case of MSSM. In section \ref{sect3}
 we use this realistic framework to discuss the contributions of ESSM to
 $(g-2)_\mu$ and to the radiative transitions $\tau\rightarrow\mu\gamma$ and
 $\mu\rightarrow e\gamma$. We will see that ESSM can naturally account for the
 indicated anomaly in $(g-2)_\mu$, but in this case vectorlike leptons and
 quarks would have to be discovered at the LHC and possibly NLC and quite
 certainly $\tau\rightarrow\mu\gamma$ and very likely also
 $\mu\rightarrow e\gamma$ should be discovered with modest improvements in the
  current limits.\\

 \section{The Yukawa Coupling Matrix in ESSM}
 \label{sect2}

 Following the discussion in the introduction (see Ref. \cite{BabuJCP} and
 \cite{JCPBabu} for details and notation), the 5$\times$5 Yukawa coupling
 matrix involving the 3 chiral ($q^i_{L,R}$) and the two vectorlike families
 ($Q_{L,R}$ and $Q'_{L,R}$) is assumed to have the simple form:
\begin{eqnarray}
\label{eq:Yuk}
\begin{array}{cc}
& \begin{array}{lcr}  q^i_L\;\;\;\; &\;\; Q_L & \;\;\;\;\;Q'_L \end{array}\\
h^{(0)}_{f,c}=\begin{array}{c} \bar{q}^i_R\\
                                 \bar{Q}_R\\
                                 \bar{Q}'_R
\end{array} &
\begin{array}({ccc}) 0_{3\times 3} & X_fH_f & Y_cH_s\\
                    Y^{'\dagger}_cH_s  & z_cH_V & 0\\
                    X^{'\dagger}_fH_f  & 0 & z'_fH_V
\end{array}~.
\end{array}
\end{eqnarray}
 Here
the symbols $q$, $Q$ and $Q'$ stand for quarks {\it as well as} leptons;
$i=1,2,3$ corresponds to the chiral families. The subscript $f$ for the
Yukawa couplings $X_f$ and $X'_f$ denotes the four sectors $u$, $d$, $l$ or
$\nu$, while $c=q$ or $l$ denotes quark or lepton color. The fields $H_f$
with $f=u,d$ denote the familiar two Higgs doublets, while $H_s$ and $H_V$
are Higgs Standard Model singlets, which can effectively be admixtures of
for example a dominant SO(10)-singlet and a sub-dominant SO(10) 45-plet
with a VEV along the B-L direction (see below). The zeros appearing 
in Eq. \eqref{eq:Yuk},
especially the direct coupling terms in the upper 3$\times$3 block, are
expected to be corrected so as to lead to masses $\lesssim$ a few $MeV$,
through VEVs inserted into higher dimensional operators. The Higgs fields
are assumed to acquire VEVs so that $\langle H_V \rangle \sim
\langle H_s\rangle \sim 1\; {\rm TeV}
 \gtrsim \langle H_u \rangle \sim 200$ GeV $\gg
\langle H_d \rangle$.

The parametrization in Eq. \eqref{eq:Yuk} anticipates that with SO(10)
 intact, even if $z_c=z'_f$, $X_f=X'_f$ and $Y_c=Y'_c$ at the GUT-scale,
 renormalization effects would introduce differences between these Yukawa
 couplings at the electroweak scale, because $Q_{L,R}$ are
 SU(2)$_L$-doublets, while $Q'_{L,R}$ are SU(2)$_L$-singlets [see Eq. (10)
 of  Ref. \cite{BabuJCP}]. Denoting $X_f^T=(x_1,x_2,x_3)_f$, and
 $Y_c^T=(y_1,y_2,y_3)_c$, it is easy to see that regardless of the values of
 these Yukawa couplings, one can always rotate the basis vectors so that
 $Y_c^T$ is transformed into $\hat{Y}_c^T=(0,0,1)y_c$, $X_{f,c}^T$
 simultaneously into the form ${\hat{X}}_f^T=(0,p_f,1)x_f$, ${X'}_f$ into
 ${\hat{X}'_f}=(0,{p'}_f,1){x'}_f$
 and ${Y'}_c$ into ${\hat{Y}'}_c=(0,0,1){y'}_c$.
 It is thus apparent why one family remains massless  (barring corrections
  $\lesssim$ a few MeV), even if there is no hierarchy in the original Yukawa
  couplings $(x_i)_f$ and $(x_i)_c$, etc., defined in the gauge basis; this
  one is naturally identified
  with the electron family. If, for simplicity, one puts $x_f=x'_f$,
  $y_c=y'_c$ and $z=z'$ at the unification scale, one would obtain
  $m^{(0)}_{t,b,\tau}\approx(2x_{f}y_c)
  (\langle H_s \rangle \langle H_f \rangle)/(z \langle H_V \rangle)$, and
  $m^{(0)}_{c,s,\mu}\approx m^{(0)}_{t,b,\tau}(p_fp'_f/4)$. Note, even if
  $p_f$ and $p'_f$ are not very small compared to unity, their product
  divided by four can still be pretty small (e.g., suppose $p_f\sim p'_f\sim
  1/2$ to $1/7$, then $p_fp'_f/4 \sim 1/16$ to $1/200$). One can thus naturally
  get a large hierarchy between the masses of the muon and the tau families
  as well, without introducing very  small numbers from the beginning.

  We stress that the parameters of the mass-matrices of the four sectors
  $u$, $d$, $l$ and $\nu$, and also those entering into $X$ versus $X'$ or
  $Y$ versus $Y'$ in a given sector, are of course not all independent,
  because a large number of them are related to each other at the GUT-scale
  by the group theory of SO(10) and the representations of the relevant Higgs
  multiplets. [For most purposes the group theory of G(224) suffices.] This
  results in an enormous reduction of parameters. [For example, with
  $(h_V {\bf 16_V} {\bf \bar{16}_V}H_V+h_{3V}{\bf 16_3}{\bf 16_V}{\bf 10_H}+
  h_{3\bar V}{\bf 16_3}{\bf \bar{16}_V}H_V)$ being the leading terms of the
  effective superpotential, one would get the following relations at the
  GUT-scale for either SO(10) or G(224): $x_u=x_d=x_l=x_\nu=x'_u=x'_d=x'_l
  =x'_\nu$; $y_q=y_l=y'_q=y'_l$ and $z_f=z'_c$. In this case, the entries
  denoted by "1" in the matrices $\hat X_f^T$, $\hat X'_f$, $\hat Y_c^T$
  and $Y'_c$ will be given by just three parameters (including $\langle H_u
  \rangle / \langle H_d\rangle=\tan\beta$) instead of sixteen - at the
  GUT-scale. Similar economy arises for the $(p_f,p'_f)$ parameters and the
  masses of the vectorlike fermions (see below).
 At this point, it is worth noting that a successful framework, based on
  MSSM embedded in SO(10), has recently been proposed \cite{BPW} that
  introduces only the three chiral families but no vectorlike families
(appropriate for MSSM) and a
  minimal Higgs system -- i.e., a single ${\bf 45_H}$, ${\bf 16_H}$,
  ${\bf \bar{16}_H}$ and ${\bf 10_H}$. Utilizing the SO(10)-invariant
  direct Yukawa couplings of the three
chiral families to each other (e.g. couplings of the type $h_{ij}
{\bf 16_i 16_j 10_H}$) it leads to  eight predictions -- including
$m^0_b\approx m^0_\tau$, $m(\nu^\tau_L)\sim 1/20$ eV, together with
$\sin^22\Theta^{osc}_{\nu_\mu\nu_\tau}\approx 0.87$-$0.96$,
$V_{ub}\approx 0.003$, $V_{us}\approx 0.22$ and $m_d\approx 8$ MeV.
Remarkably enough,
all of these are in agreement with observation to within 10 \%. It is
interesting, as we show below, that the same form of effective mass matrix
of the three chiral (especially $\mu$ and $\tau$) families can also be
obtained for the case of ESSM simply by imposing an SO(10) group structure
analogous to that of Ref. \cite{BPW} on the off-diagonal Yukawa couplings
of ESSM [shown in Eq. \eqref{eq:Yuk}], and by performing a see-saw block
diagonalization that integrates out the heavy vectorlike families. Thus,
the successes of Ref. \cite{BPW} can essentially be retained for the case
of ESSM, embedded in SO(10), as well. To see this briefly (details will be
given in a separate note), let us go to the basis, denoted by a hat as above,
in which the first family is entirely (or almost) decoupled from the two
vectorlike families, so that $x_1=x'_1= y_1=y'_1=y_2=y'_2=0$
\cite{NewFN}. For the
convenience of writing, we drop the hat on the Yukawa couplings. Using only
(${\bf 10_H}$, ${\bf 16_H}$, ${\bf \bar{16}_H}$ and ${\bf 45_H}$) the
relevant leading terms of the effective superpotential involving the two
chiral (${\bf 16_2}$ and ${\bf 16_3}$) and the two vectorlike families
(${\bf 16_V}+{\bf \bar{16}_V}$) that would conform with the Yukawa coupling
matrix of Eq. \eqref{eq:Yuk} and also would yield (after integrating out
$Q$ and $Q'$) a mass matrix analogous to that of Ref. \cite{BPW} is
given by:
\begin{eqnarray}
\label{eq:WYuk}
\hat{W}_{Yuk}&=&h_V{\bf 16_V \bar{16}_V} H_V +
f_V{\bf 16_V \bar{16}_V (45_{H}}/M){\bf 1'_V}+
h_{3V}{\bf 16_3 16_V 10_H}\nonumber\\
&+&\tilde{h}_{3V}{\bf 16_3 16_V 10_H 45_{H}}/M+
h_{3\bar{V}}{\bf 16_3 \bar{16}_V} H_s+
h_{2V}{\bf 16_2 16_V 10_H}\Big(\frac{X}{M}\Big)\\
&+&a_{2V}{\bf 16_2 16_V 10_H 45_{H}}/M+
g_{2V}{\bf 16_2 16_V 16_H^d 16_{H}}/M\nonumber
\end{eqnarray}
where $v_0\equiv\langle H_V\rangle\sim{\bf 1'_V}\sim 1$ TeV $\sim
 \langle H_s\rangle>\langle H_u\rangle\sim 200\; GeV \gg \langle H_d\rangle$;
$\langle {\bf 16_H}\rangle\sim \langle {\bf 45_H}\rangle\sim
\langle X \rangle\sim M_{GUT}$ and $M\sim M_{string}$. It is presumed that
owing to flavor symmetries \cite{Flavor}
, $f_V$ and $h_{2V}$ terms require the presence
of ${\bf 45_H}$ and $X$, respectively, so that they are suppressed by one
power of $\langle {\bf 45_H}\rangle/M$ or $\langle X/M \rangle\sim
(1/3$-$1/10)$ compared to the $h_{3V}$ term \cite{FN1}. The $\tilde{h}_{3V}$,
$a_{2V}$ and $g_{2V}$ terms are also naturally suppressed (by SO(10) group
theory) by a similar factor relative to the $h_{3V}$ and $h_{V}$ terms. Note,
the VEV $\langle {\bf 45_H}\rangle \propto B-L$ introduces a
$B-L$ dependence, while $\langle {\bf 16_H^d}\rangle$ introduces up-down
distinction [here $\langle {\bf 16_H^d}\rangle$ denotes the electroweak VEV
of ${\bf 16_H}$, which arises through a mixing between ${\bf 16_H}$ and
${\bf 10_H^d}$; see Ref. \cite{BPW}]. Taking these into account, the
parameters (or corresponding VEVs of certain entries) of the Yukawa matrix
\eqref{eq:Yuk} in the rotated hat basis (discussed above)
are given by (hat is suppressed):
\begin{eqnarray}
\label{eq:YukGUT}
M_Q&=& z_c\langle H_V\rangle=h_V^Qv_0(1+\kappa_{B-L})\nonumber\\
M_{Q'}&=& z'_f\langle H_V\rangle=h_V^{Q'}v_0(1-\kappa_{B-L})\nonumber\\
x_f&=& h_{3V}^Q(1+\delta_{B-L})\nonumber\\
x'_f&=& h_{3V}^{Q'}(1-\delta_{B-L})\\
p_fx_f&\equiv &x_{2f}=h^Q_{2V}\Big\langle\frac{X}{M}\Big\rangle(1+\xi_{B-L})
+[ g^{Q}_{2V} \sin \gamma \langle{\bf 16_H}\rangle/M]_d\nonumber\\
p'_fx'_f&\equiv &x'_{2f}=h^{Q'}_{2V}\Big\langle\frac{X}{M}\Big\rangle
(1-\xi_{B-L})
+[g^{Q'}_{2V}\sin\gamma\langle {\bf 16_H}\rangle/M]_d\nonumber\\
y^Q&=&y^{\prime Q'}=h_{3\bar{V}}\nonumber
\end{eqnarray}
These entries correspond to GUT scale values.  
The superscripts $Q$ and $Q'$ on $h_V$ (and likewise on the other couplings)
signify that even if $h^Q_{V}=h^{Q'}_{V}$ at the GUT-scale (owing to SO(10)),
renormalization effects would introduce differences between the two couplings
at the electroweak scale [see Eq. (10), Ref. \cite{BabuJCP}]. Here,
$\kappa_{B-L}=\kappa$ for the heavy quarks ($U$, $D$, $U'$ and $D'$), while
$\kappa_{B-L}=-3\kappa$ for the heavy leptons ($E$ and $E'$); likewise
$\delta_{B-L}\equiv(\delta, -3\delta)$ for $(q,l)$, and
$\xi_{B-L}\equiv(\xi, -3\xi)$ for $(q,l)$. The second term in $x_{2f}$ and
$x'_{2f}$ contributes only to the down quarks and charged leptons. The
parameter $\sin\gamma$ denotes the mixing between $\langle {\bf 16_H}\rangle$
and $\langle {\bf 10_d}\rangle$, where
$\cos \gamma\approx (m_b/m_t)\tan \beta$ (see Ref. \cite{BPW}). Since
$\langle {\bf 45_H}\rangle/M$ is expected to be small compared to unity,
we expect the $(B-L)$-dependent parameters $\kappa$ and $\delta$
to be typically $\leq 1/10$; however, with $a_{2V}\sim g_{2V}$ and
$\langle X/M\rangle \sim \langle {\bf 45_H}\rangle/M$,
$\xi_{B-L}$ (if present) is
expected to be of the order of unity. It turns out that with the Yukawa
couplings presented in Eqs. \eqref{eq:Yuk} and \eqref{eq:YukGUT}, together
with the much suppressed direct Yukawa couplings of the electron with the
muon and the tau families, all the successes of Ref. \cite{BPW} are
essentially preserved.  This can be seen by integrating out the vectorlike
families and examining the resulting $3 \times 3$ matrix for the light
chiral families, which will have the same form as the mass matrices
of Ref. \cite{BPW} (in the leading see-saw approximation).  There is
one difference however in the prediction for $m_b$.  Owing to renormalization
effects corresponding to the running of the scale from $M_{GUT}$ to
$M_S\sim 1$ TeV, which distinguish between $M_D$, $M_{D'}$, $M_E$ and
$M_{E'}$ (see Ref. \cite{BabuJCP}), the ratio $m_b/m_\tau$ evaluated at $M_S$
for ESSM (with $\kappa=\delta=0$) turns out to be typically
larger than that for MSSM
\cite{FN2} by nearly 20-25\%. The $(B-L)$ dependent entries $\kappa$ and 
$\delta$
exhibited in Eq. \eqref{eq:YukGUT}, which are expected to be of order 1/10,
would however have the right magnitude and the right sign (if $\kappa\delta$ 
is negative) to compensate adequately for this difference. In short, the
pattern of Yukawa couplings given by Eq. \eqref{eq:Yuk}, \eqref{eq:WYuk} and
\eqref{eq:YukGUT} does correspond to a {\it realistic mass matrix} for
fermion masses and mixings in the case of ESSM, which preserves the major
successes of Ref. \cite{BPW} including especially the predictions of
$m(\nu_\tau)$, $\nu_\mu$-$\nu_\tau$ oscillation angle, $V_{cb}$ and $V_{ub}$.

Details of the analysis of the fermion masses and mixings for the case of
ESSM embedded in SO(10), including the $\kappa$ and $\delta$-terms, will be
presented in a separate paper. Here, our main focus will be to study the
new contributions to radiative transitions in the charged lepton sector
that arise for the case of ESSM.

As we shall see shortly the new contribution to the amplitude for
$\tau\rightarrow\mu\gamma$ arising from $H_d$-$H_s$ mixing would vanish
if $m_E=m_{E'}$. We note, however, that even if $E$ and $E'$ were exactly
degenerate at the GUT-scale [i.e., with $f_V=0$ and thus $\kappa=0$, see
Eqs.
\eqref{eq:WYuk} and \eqref{eq:YukGUT}], renormalization effects would split
them near the electroweak scale, because $E$ couples with $W_L$, but
$E'$ does not. For instance, with $M_E=M_{E'}$ and $h_V$ large
($\approx$ 1-2, say), at GUT scale, one finds
$(M_E/M_{E'})_{1\; TeV}=(z_l/z'_l)_{1\; TeV}\approx 0.273/0.185\approx
1.47$ [see Eq. (10), Ref. \cite{BabuJCP}]. In the presence of the 
$\kappa$-term,
which seems to be needed to account for the observed value of $(m_b/m_\tau)$
(see remarks above), it thus seems quite plausible (with $\kappa >0$) that 
the
degree of non-degeneracy of $E$ and $E'$ at the electroweak scale could
lie typically in the range of (10-50)\% (say) \cite{FN3}. Thus, for
concreteness in our analysys, that would be relevant especially for
consideration of $\tau\rightarrow\mu\gamma$, we would allow:
\begin{eqnarray}
\label{eq:MEME}
(M_E/M_{E'})_{1\; TeV}\approx 1+(10 \mbox{ to } 50)\; \%~.
\end{eqnarray}
Now see-saw diagonalization of the 5$\times$5 mass-matrix for charged
leptons, following from Eqs. \eqref{eq:Yuk} and \eqref{eq:YukGUT}, leads
to a $\mu$-$\tau$ mass-matrix given by:
\begin{eqnarray}
\label{eq:mutauMass}
\begin{array}{cc}
& \begin{array}{lr}  \mu_L\qquad\qquad\qquad & \quad\tau_L\qquad\qquad 
\;\;\;\end{array}\\
M_{\mu\tau}=\begin{array}{c} \bar{\mu}_R\\
                             \bar{\tau}_R\\
\end{array} &
\begin{array}({cc}) 0 & xy'p/M_E\\
                    x'yp'/M_{E'} & xy'/M_E+x'y/M_{E'}
\end{array}v_dv_s
\end{array}~.
\end{eqnarray}
Here, all the entries ($x$, $x'$, $y$, $y'$, $p$ and $p'$) refer to the
charged lepton sector (so the subscript $l$ is suppressed). Using the
parameters
appearing in $x_2$, $x'_2$, $x$ and $x'$ in Eq. \eqref{eq:YukGUT}, and
anticipating a correspondence with Ref. \cite{BPW}, one can express $p_l$
and $p'_l$ for charged leptons in terms of two effective parameters -- i.e.,
\begin{eqnarray}
\label{eq:pp}
p\equiv 2(\eta+3\epsilon);\qquad\qquad p'\equiv 2(\eta-3\epsilon)~.
\end{eqnarray}
Owing to the SO(10)-constraint, the corresponding parameters $p$ and $p'$
for the $b$-$s$ sector are $p_d=2(\eta-\epsilon)$ and $p'_d=2(\eta+\epsilon)$,
respectively
(compare with Ref. \cite{BPW}). Note, if we drop the relatively small
$(B-L)$-dependent $\delta$ and $\kappa$ terms [$\cal O$(1/10)] in $x$ ,$x'$,
$z$ and $z'$;
see Eq. \eqref{eq:YukGUT}, we would have $x=x'$, $y=y'$ and $z=z'$
(therefore $M_E=M_{E'}$) at the GUT-scale due to SO(10) \cite{FN4}.
In this case, using $M_E=zv_0$ and $M_{E'}=z'v_0$, one would have the
equality of the ratios:
\begin{eqnarray}
\label{eq:ratios}
xy'/M_E=x'y/M_{E'}
\end{eqnarray}
at the GUT scale. This would lead to {\it a very simple form} for the
$\mu$-$\tau$ mass-matrix [see Eq. \eqref{eq:mutauMass}], with $(xy'/M_E)$
being a common factor in all three elements of the matrix.

It is worth noting that in the context of the renormalization effects studied
in Ref. \cite{BabuJCP} [where it was assumed that all the Yukawa couplings
of vectorlike and the third family of fermions -- $x_f$, $x'_f$, $y_c$,
$y'_c$, $z_f$ and $z'_c$ -- are large ($\sim$ 1 to 2) at the GUT-scale so
that they acquire their respective quasi-fixed point values at the electroweak
scale], the equality \eqref{eq:ratios} and thus the simple form of the
mass-matrix referred to above holds even at the electroweak scale. This is
because, the ratio of the renormalized couplings at the electroweak scale --
for instance for the leptons [see Eq. (10), Ref. \cite{BabuJCP}] -- given by
$xy'/z\approx 0.396\times 0.251/0.273 \approx 0.364$ equals the
ratio $x'y/z'\approx 0.368\times 0.184/0.185 \approx 0.364$. Analogous
equalities for the renormalized couplings are found to hold
(see Ref. \cite{BabuJCP}) for the quark sector as well.

As a further remark, as long as $x=x'$, $y=y'$ and $z=z'$ at the GUT-scale
(i.e., in the limit $\kappa=\delta=0$), we would in fact expect the equality
\eqref{eq:ratios} to hold at the EW scale to a fairly good approximation
(better than 10\%), even if not all the Yukawa couplings are so large at
the GUT-scale as to approach their quasi-fixed point values at the electroweak 
scale.

In the interest of simplicity in writing analytic expressions for the mixing
angles, which would be relevant to radiative transitions, we would ignore the
$(B-L)$-dependent $\delta$ and $\kappa$ terms which are ${\cal O}(1/10)$ and
assume (for reasons explained above) that the
equality \eqref{eq:ratios} and thus the simple form of the $\mu$-$\tau$
mass-matrix holds to a good approximation at the electroweak scale. This
would amount to making an error typically of 10-25 \% in the radiative
amplitudes \cite{FN5}, which would, however, not affect our conclusion.
A more refined analysis will be presented elsewhere.

With the equality \eqref{eq:ratios} holding (approximately) at the electroweak
scale and the corresponding simple form of the
mass-matrix, that results from Eq. \eqref{eq:mutauMass},
one can identify the parameters $\eta$ and $\epsilon$ appearing in
Eq. \eqref{eq:pp} precisely with those in Ref. \cite{BPW}. From the
fitting of fermion masses carried out in Ref. \cite{BPW}, one then has:
$\eta\approx -0.15$ and $\epsilon\approx 0.095$, and thus
[see Eq. \eqref{eq:pp}]:
\begin{eqnarray}
\label{eq:ppvalues}
p_l\approx 0.27;\qquad\qquad p'_l\approx -0.87~.
\end{eqnarray}
The $\mu$-$\tau$ mass-matrix \eqref{eq:mutauMass}, subject to
Eq. \eqref{eq:ratios}, gets diagonalized by the simple 2$\times$2 matrices:
\begin{eqnarray}
\label{eq:UU}
U_L\approx\left[\begin{array}{cc} 1 & p'_l/2 \\
                                -p'_l/2 & 1
\end{array} \right];\qquad\qquad
U_R\approx\left[\begin{array}{cc} 1 & p_l/2 \\
                                -p_l/2 & 1
\end{array} \right]
\end{eqnarray}
and one gets:
\begin{eqnarray}
\label{eq:MmuMtau}
m_\mu\approx -\left(\frac{p_lp'_l}{2}\right)\left(\frac{xy'}{M_E}\right)_l
v_dv_s;\qquad\qquad
m_\tau\approx \left(2+\frac{p_lp'_l}{2}\right)\left(\frac{xy'}{M_E}\right)_l
v_dv_s~.
\end{eqnarray}
Thus, $m_\mu/m_\tau\approx-p_lp'_l/4\approx 1/17$, in good accord with
observation. Analogous discussion will apply to the quark and neutrino
sectors, which are not relevant here.

\section{Radiative Transitions in ESSM}
\label{sect3}

We will be interested in radiative transitions of charged leptons, in
particular $\tau\rightarrow \mu\gamma$, $\mu\rightarrow \mu\gamma$ and
$\mu\rightarrow e\gamma$. The corresponding amplitudes are defined by:
\begin{eqnarray}
\label{eq:amplitude}
A(\Psi^i_L \rightarrow \Psi^j_R \gamma)\equiv A_{ij}
 (\bar{\Psi}_{jR}\sigma_{\mu\nu}q^\nu \Psi_{iL})\epsilon^\mu~.
\end{eqnarray}
New contributions to these amplitudes would arise from (a) scalar loops
involving $(H_d, H_s)$ mixing (see Figs 1a,b), and also (b) (W and Z)-loops
(see later). These new contributions arise from the Yukawa couplings of
the chiral with the vectorlike families (i.e., $x_i$, $y_i$, $x'_i$ and
$y'_i$) [see Eq. \eqref{eq:Yuk}]. We will use the hat basis discussed above,
although not exhibit the hats (thus $x_1=x'_1=y_1=y'_1=y_2=y'_2=0$). We
drop the subscripts $f$ and $c$, both of which now correspond to charged
leptons. The scalar loops are evaluated by exchanging the mass eigenstates
$H_1=H_d\cos\theta + H_s\sin\theta$, and $H_2=-H_d\sin\theta + H_s\cos\theta$,
with eigenvalues $M_1$ and $M_2$, respectively. The amplitude is found to be:
\begin{eqnarray}
\label{eq:amplitude2}
A_{ij}^{H_dH_s}&=&(y'_ix_j/m_E){\cal K}_E+(x'_iy_j/m_{E'}){\cal K}_{E'}
\nonumber\\
&=&\left(\frac{y'_ix_j}{m_E}+\frac{x'_iy_j}{m_{E'}}\right)
\frac{{\cal K}_E+{\cal K}_{E'}}{2}+
\left(\frac{y'_ix_j}{m_E}-\frac{x'_iy_j}{m_{E'}}\right)\frac{{\cal K}_E-
{\cal K}_{E'}}{2}\\
&\equiv& A_{ij}^{(+)}+A_{ij}^{(-)}\nonumber
\end{eqnarray}
where ${\cal K}_F$ ($F=E$ or $E'$) is given by \cite{NathMarciano}:
\begin{eqnarray}
\label{eq:KF}
{\cal K}_F=-\frac{e\sin\theta\cos\theta\;(M_F)}{32\pi^2}
[{\cal B}(M_F, M_1)- {\cal B}(M_F, M_2)]
\end{eqnarray}
with,
\begin{eqnarray}
\label{eq:MF}
M_F {\cal B}(M_F, M_i)=\frac{r_i}{(1-r_i)^2}\left[3-r_i+\frac{2\ln r_i}
{1-r_i}\right]
\end{eqnarray}
where
\begin{eqnarray}
\label{eq:ri}
r_i=M_F^2/M_i^2~.
\end{eqnarray}
Now $H_d$-$H_s$ mixing denoted by the angle $\theta$ can arise
through (i) a term in the superpotential $W \supset \lambda
H_uH_dH_s$ \cite{FN6}, and (ii) a soft SUSY-breaking term $AH_uH_dH_s$
(involving only scalar fields). Using $\langle H_d^{\dagger}\rangle=v_d$
and $\langle H_s^{\dagger}\rangle=v_s$, these two terms together would induce
a mass-mixing term $[(\hat{\lambda}v_dv_s)(H_dH_s)+h.c.]$ where
\begin{eqnarray}
\label{eq:hatlambda}
\hat{\lambda}=\lambda^2+(A/v_s)\tan\beta.
\end{eqnarray}
Correspondingly, one obtains, for the $H_d-H_s$ mixing angle:
\begin{eqnarray}
\label{eq:theta}
\sin\theta\cos\theta=(\hat{\lambda}v_dv_s)/(M_2^2-M_1^2)
\end{eqnarray}
We should expect $\hat{\lambda}$ to be complex in general, owing to
the phases in the $A$--term and/or $v_s$, but for now we shall assume
$\hat{\lambda}$ to be real.  We will comment in subsection 4.2
on the implications of
a complex $\hat{\lambda}$ on the EDM of the muon, which turns out to be
in the observable range in proposed experiments.  
At this stage, it is worth noting that to leading order in see-saw
diagonalization, which serves to integrate out the heavy vectorlike
families $(Q, Q')$, the mass matrix of the charged leptons in the three
chiral families (baring small corrections $\lesssim$ few $MeV$ that arise
from direct entries in the 3$\times$3 block of Eq. \eqref{eq:Yuk}, see
discussions above) are given by:
\begin{eqnarray}
\label{eq:Mij}
M_{ij}=\left(\frac{y'_ix_j}{m_E}+\frac{x'_iy_j}{m_{E'}}\right)v_dv_s~.
\end{eqnarray}
Now, for discussions of $(g-2)_\mu$ and $\tau\rightarrow\mu\gamma$, we may
ignore the electron family; thus $M_{ij}$ is effectively a 2$\times$2
matrix, and so is $A_{ij}$. Note that $A^{(+)}_{ij}$ of
Eq. \eqref{eq:amplitude2} is directly proportional to the mass-matrix
$M_{ij}$. As a result, as we go to the physical basis by diagonalizing
$M_{ij}$, $A^{(+)}_{ij}$ gets diagonalized as well. {\it Thus, to a very good
approximation,} $A^{(+)}_{ij}$ {\it does not contribute to off-diagonal
transitions like} $\tau\rightarrow\mu\gamma$ (likewise, the analogous term
in the quark sector does not contribute to $b \rightarrow s\gamma$){\it, but}
$A^{(-)}_{ij}$ {\it does}. On the other hand, $A^{(+)}_{ij}$ makes bigger
contribution, compared to $A^{(-)}_{ij}$, to diagonal transitions -- that is
to $(g-2)$ of the muon and the tau. We see from Eqs. \eqref{eq:amplitude2}
and \eqref{eq:KF} that ${\cal K}_F$ and therefore the new contributions to
$\tau \rightarrow \mu \gamma$
arising from Fig. 1 would tend to vanish
if  $M_E\rightarrow M_{E'}$ (because in this case,
 ${\cal K}_E\rightarrow {\cal K}_{E'}$ and thus $A^{(-)}_{ij}\rightarrow 0$). 
While we
 expect $M_E\sim M_{E'}$, we do not of course have any
 reason to expect exact degeneracy of  $E$ and
 $E'$. For numerical purposes, we would take $M_1$ and $M_2$ to be
 comparable to within a factor of two ($M_1$ by choice being lighter)
 and $(M_E/M_{E'})$ to be away from unity as in Eq. \eqref{eq:MEME}.

 To evaluate the new contributions to radiative transitions, we first go to
 the physical basis by diagonalizing the $\mu$-$\tau$ mass-matrix $M_{ij}$
 with the transformation $M\rightarrow\hat{M}=U_R^{\dagger}MU_L$
 [see Eqs \eqref{eq:mutauMass}, \eqref{eq:ratios} and \eqref{eq:UU}], and
 then impose the same transformation on the matrix $A_{ij}$
 [Eq. \eqref{eq:amplitude2}]; so that $A\rightarrow\hat{A}=U_R^{\dagger}AU_L$.
 The matrices $U_{L,R}$ are given approximately by Eq. \eqref{eq:UU}.
Noting that the diagonal elements of $\hat{M}$ are just $m_\mu$ and $m_\tau$,
which are proportional to those of $\hat{A}^{(+)}$, one then
straightforwardly obtains:
\begin{eqnarray}
\label{eq:amu}
a_\mu^H\equiv
a_\mu^{H_dH_s}\approx \frac{m_\mu^2}{e}({\cal K}_E+{\cal K}_{E'})/(v_dv_s)
\end{eqnarray}
where $m_\mu$ stands for $(-p_lp'_l/4)m_\tau$ [see Eq. \eqref{eq:MmuMtau}].
Likewise, using contribution from $\hat{A}^{(-)}$
[see Eq. \eqref{eq:amplitude2}], one obtains:
\begin{eqnarray}
\label{eq:ALR}
A_L^H\equiv
A(\tau_L\rightarrow\mu_R)^{H_dH_s}&\approx& p(1+p^{'2}/4)(x_3y'_3/M_E)
({\cal K}_E-{\cal K}_{E'})/2\nonumber\\
&\approx& \left\{p(1+p^{'2}/4)\frac{m_\tau}{2+pp'/2}\right\}
\frac{1}{v_dv_s}({\cal K}_E-{\cal K}_{E'})/2\\
\label{eq:ARL}
A_R^H\equiv
A(\tau_R\rightarrow\mu_L)^{H_dH_s}&\approx& -p'(1+p^2/4)(x'_3y_3/M_{E'})
({\cal K}_E-{\cal K}_{E'})/2\nonumber\\
&\approx& \left\{-p'(1+p^2/4)\frac{m_\tau}{2+pp'/2}\right\}
\frac{1}{v_dv_s}({\cal K}_E-{\cal K}_{E'})/2
\end{eqnarray}
where we have used Eq. \eqref{eq:MmuMtau} for $m_\tau$.
All the parameters $p$, $p'$, etc., correspond to the charged lepton sector.
It thus follows that
for a given choice of the spectrum ($M_E$, $M_{E'}$, $M_1$, $M_2$) and
$\hat{\lambda}$ [see Eq \eqref{eq:hatlambda}], we can calculate $a_\mu$ and
$A(\tau_{L,R}\rightarrow\mu_{L,R}+\gamma)$ arising from $H_d$-$H_s$ mixing.
Note
that $v_dv_s$ appearing in the denominator in Eqs. \eqref{eq:ALR} and
\eqref{eq:ARL} cancels out because ${\cal K}_F\propto v_dv_s$
[see Eqs \eqref{eq:KF} and \eqref{eq:theta}]. We now proceed with the
numerical evaluation of these radiative amplitudes for a few sample choices
of the spectrum. They are listed in Table 1. We will return to a discussion
of these contributions after presenting the contributions from the $W$-loop.

It should be mentioned that the supersymmetric partners of $(E_{L,R},
E'_{L,R})$ will also contribute to radiative transitions in the lepton
sector.  Such contributions will arise through diagrams analogous to
Fig. 1, obtained by replacing ($H_s,~H_d$) fields by their fermionic partners
$(\tilde{H}_s,~\tilde{H}_d$), $E_{L,R}$  by the scalar heavy
leptons $\tilde{E}_{L,R}$ and $E'_{L,R}$ by $\tilde{E'}_{L,R}$.  These
diagrams are however suppressed somewhat relative to those shown in
Fig. 1, mainly because the masses of the vector sleptons ($\tilde{E},
~\tilde{E'}$) are expected to be much larger than those of the fermions
($E,~E'$).  The scalar heavy leptons receive masses from the superpotential
as well as from the soft SUSY breaking terms.  For example, if
$M_E = 300$ GeV and $m_0 = 500$ GeV (the soft SUSY breaking scalar mass
parameter), then $M_{\tilde{E}} \simeq (M_E^2 + m_0^2)^{1/2} \simeq 580$
GeV, to be compared with the masses $M_{1,2}$ of the Higgs fields $H_d$
and $H_s$ of Fig. 1 which are in the range $100-250$ GeV.  (The 
$\tilde{H}_d-\tilde{H}_s$ Higgsino mass term is comparable to $M_E$.)
In any case, the flavor structure of these supersymmetric diagrams are 
identical to those in Fig. 1, so even if the new diagrams have
comparable magnitudes, their effects can be mimicked by a redefinition
of $\hat{\lambda}$.  Thus we shall focus on the diagrams of Fig. 1
in our numerical evaluation of the radiative transitions.  

\subsection{New Contributions from the $W$-Loop}
\label{sect4}
In ESSM, both $(N_L, E_L)$ and $(N_R, E_R)$ are doublets of SU(2)$_L$; thus
they both couple to $W_L$, while $(N'_L, E'_L)$ and $(N'_R,E'_R)$ do not.
We will argue that the new (non-standard) contributions from the $W$-loop are
strongly suppressed compared to those from the $H_d$-$H_s$ loop.
Allowing for the mass-mixing of the light and the
heavy leptons [see Eq. \eqref{eq:Yuk}], the weak interaction Lagrangian
contains terms given by:
\begin{eqnarray}
\label{eq:lagrangeian}
{\cal L}_W^{(N)}=
\left(g_W/\sqrt{2}\right)\sum_{i,a}\left[\bar{N}_L^a\gamma_\mu V_{iN_L^a}
\Psi^i_L+
\bar{N}_R^a\gamma_\mu V_{iN_R^a}\Psi^i_R\right]W^\mu_L+h.c.
\end{eqnarray}
Here $\Psi_{iL,R}$ denote {\it physical} charged leptons
$(\mu,\tau,E)_{L,R}$, and $(N_{L,R}^a)_{a=1,2}$ denote the physical neutral
heavy leptons given by
\begin{eqnarray}
N^1_{L,R}&=&\cos\Theta_{L,R}^NN_{L,R}+\sin\Theta_{L,R}^NN'_{L,R}\\
N^2_{L,R}&=&-\sin\Theta_{L,R}^NN_{L,R}+\cos\Theta_{L,R}^NN'_{L,R}\nonumber
\end{eqnarray}
Note that these include $N$-$N'$ mixing which is induced by the mass-matrix
of Eq. \eqref{eq:Yuk}. We refer the reader to  Ref. \cite{PS95}
for diagonalization
of the $Q$-$Q'$ mass matrices in all four sectors and for expressions of the
mixing angles. It is argued there that, including renormaliztion group
effects, the mixing angles $\Theta_L^N$ and $\Theta_R^N$ are nearly equal
(to better than 5\%). The
coefficients $V_{iN_L^a}$ and $V_{iN_R^a}$ are obtained by diagonalizing the
mass-matrices in the leptonic up and down sectors (analogous to the
CKM-matrix). The new contributions to radiative transitions due to the
$W$-loop are shown in Fig. 2. Using Ref. \cite{BabuEcker}, the contribution
of the $W$-loop to the radiative amplitude $A_{ij}$
[defined in Eq. \eqref{eq:amplitude}] is given by:
\begin{eqnarray}
\label{eq:amplitude3}
A^{(W)}_{ij}=\frac{eM_N}{32\pi^2}\left(g_W^2/m_W^2\right)
\sum_{a=1,2}\left(V_{iN_L^a}V_{jN_R^a}^*\right)F(x)
\end{eqnarray}
\begin{eqnarray}
where
\label{eq:F}
F(x)= (2-\frac{15}{2}x+6x^2-3x^2\ln x-\frac{x^3}{2})/(1-x)^3
\end{eqnarray}
Here $x\equiv M_N^2/m_W^2$ and the quantities $m^2_{\mu,\tau}/M_N^2$ are
dropped. Diagonalizing the mass-matrix for the
charged and neutral leptons, we get \cite{PS95}:
\begin{eqnarray}
\label{eq:VVVV}
V_{\tau_LN_L^1}&\approx& -\left(\frac{\kappa_u^2}{\kappa_\lambda^2}\right)
\left(\frac{\kappa_s}{\kappa_\lambda}\right)
 \left(\frac{1}{\eta_L^2-1}\right)\nonumber\\
V_{\tau_LN_L^2}&\approx& \left(\frac{\kappa_u}{\kappa_\lambda}\right)
 \left[1-\frac{\kappa_r^2}{\kappa_\lambda^2}+\frac{3}{8}
 \frac{\kappa_r^4}{\kappa_\lambda^4}\right]\\
 V_{\tau_RN_R^1}&\approx&\frac{\kappa_d}{\kappa_\lambda}\nonumber\\
V_{\tau_RN_R^2}&\approx& -\left(\frac{\kappa_u}{\kappa_\lambda}\right)
\left(\frac{\kappa_s}{\kappa_\lambda}\right)
\left(\frac{\kappa_d}{\kappa_\lambda}\right)
 \left(\frac{\eta_L}{\eta_L^2-1}\right)~. \nonumber 
\end{eqnarray}
Following the notations of Ref. \cite{PS95} and \cite{BabuJCP},
$\kappa_u\equiv x_3\langle H_u \rangle=x_3 v_u$,
$\kappa_d\equiv x_3\langle H_d \rangle=x_3 v_d$,
$\kappa_\lambda\equiv z\langle H_v \rangle=z v_0\approx M_N\approx M_E$
(putting $z_f=z_c=z$ at the GUT-scale), and
$\kappa_s\equiv y\langle H_s \rangle=yv_s$. The entity $\eta_L$ denotes
the renormalization of the Yukawa couplings due to SU(2)$_L$ gauge
interactions for the effective momentum running from $M_{\rm GUT}$ to the
electroweak scale ($\eta_L\approx 1.5$, see Ref. \cite{PS95}). In writing
Eq. \eqref{eq:amplitude3}, we have made the approximation that
$\kappa_u\ll \kappa_\lambda$ and $\kappa_s < \kappa_\lambda$, and neglected
 the relevant small terms. For $M_N\sim 500$ GeV, we expect
$\eta_u\equiv \kappa_u/\kappa_\lambda\approx 1/5$-1/20 (see footnote [27]
in Ref. \cite{BP2002}); in particular, an explanation of the possible
NuTeV-anomaly (if it is real) suggests $\eta_u\approx 1/10$-1/15. For the
estimate presented below, we would use:
$\eta_u\equiv \kappa_u/\kappa_\lambda\approx 1/10$.

The vertices given above for $W^+_L\rightarrow\tau_LN^a_L$ and
$W^+_L\rightarrow\tau_RN^a_R$ would give the corresponding vertices for
 $W^+_L\rightarrow\mu_LN^a_L$ and $W^+_L\rightarrow\mu_RN^a_R$, with the
 insertion of an additional factor of $(p'_l/2)$ for the substitution
 $\tau_L\rightarrow\mu_L$ and of $(p_l/2)$ for $\tau_R\rightarrow\mu_R$.
 Using these substitutions and Eq. \eqref{eq:VVVV}, the sum of the
 contributions from the $N_1$ and $N_2$-lines in the loop (Fig. 2) is given
 by :
 \begin{eqnarray}
\label{eq:A1}
A_{N_1+N_2}^W(\tau_L\rightarrow\mu_R\gamma)\approx e
\left(\frac{\alpha_2}{8\pi}\right)
\left(\frac{M_NF(x)}{m^2_W}\right)
\left(\frac{\eta_u^2}{\eta_L+1}\right)
\left(\frac{m_\tau}{2M_N}\right)
\left(\frac{p_l}{2}\right)\\
\label{eq:A2}
 A_{N_1+N_2}^W(\tau_R\rightarrow\mu_L\gamma)\approx e
\left(\frac{\alpha_2}{8\pi}\right)
\left(\frac{M_NF(x)}{m^2_W}\right)
\left(\frac{\eta_u^2}{\eta_L+1}\right)
\left(\frac{-m_\tau}{2M_N}\right)
\left(\frac{p'_l}{2}\right)~.
 \end{eqnarray}
 In above, we have used $(\kappa_u\kappa_s/\kappa_\lambda)\approx m_\tau/2$
 -- [see Eq. \eqref{eq:MmuMtau}]. Evaluating the functions $F(x)$, $\kappa_E$
 and $\kappa_{E'}$ numerically, we find that {\it because of the suppression
factor $\eta_u^2\approx 10^{-2}$} in Eqs. \eqref{eq:A1} and \eqref{eq:A2},
the W-contributions to $a_\mu$ and to the $\tau\rightarrow\mu\gamma$
amplitude are strongly suppressed compared to those of the $H_d$-$H_s$ loop,
as long as $\hat\lambda \geq 5$. To be specific, we obtain: 
$(a_\mu^W/a_\mu^{H_d-H_s} \leq 1/50$, and 
$\left|A^W(\tau \rightarrow \mu \gamma)/A^{H_d-H_s}(\tau \rightarrow 
\mu \gamma)\right|
\leq 1/10$ for $\hat{\lambda} \geq 5$.  
Of course, if $\hat\lambda$ is substantially
less than 5, both the $W$ and the $H_d$-$H_s$ loop contributions to $a_\mu$ 
as well as to
$A(\tau\rightarrow\mu\gamma)$ would be comparable, but rather small.
Henceforth, we will use $\hat\lambda \geq 4$ (which is quite plausible
for $\tan\beta\geq 3$-5 (say)), and drop the $W$-loop contribution to
$(g-2)_\mu$ and to the $\tau\rightarrow\mu\gamma$-amplitude.
One can verify that the non-standard $Z^0$-loop contributions
involving $E$ and $E'$ in the loop
are extremely small ($\lesssim$ 1 \%) compared to those from the
$W$-loop \cite{FN34}. They are therefore dropped as well in subsequent
discussions.
The contributions from the $H_d$-$H_s$ loop are listed in Table
1. The rate for $\tau\rightarrow\mu\gamma$ is calculated by using:
\begin{eqnarray}
\label{eq:taumugamma}
\Gamma ( \tau\rightarrow\mu\gamma )=\left[ |A(\tau_L\rightarrow\mu_R )|^2+
|A(\tau_R\rightarrow\mu_L )|^2\right] m_\tau^3/(16\pi )
\end{eqnarray}
where the amplitudes defined by Eq. \eqref{eq:amplitude} include
contributions from only the $H_d$-$H_s$ loop.\\

\noindent\begin{tabular}{|c|c|c|c|c|}
\hline
$(M_1,M_2,M_E,M_{E'})$ & $\hat\lambda$ &
$(g-2)_\mu\times 10^{10}$ & $A(\tau\rightarrow\mu\gamma)$ &
BR$(\tau\rightarrow\mu\gamma)$\\
&&$a_\mu^H$&
$(A_L^H,A_R^H)\times 10^9$ GeV&
\\ \hline
(1) (120, 200, 320, 280)&10&29.6&(0.79, 2.18)&2.7$\times 10^{-7}$\\ \hline
(2) (120, 200, 320, 280)&4&11.8&(0.32, 0.87)&4.3$\times 10^{-8}$\\ \hline
(3) (120, 200, 320, 220)&10&33&(2.2, 6.06)&2.1$\times 10^{-6}$\\ \hline
(4) (120, 250, 420, 300)&12.5&26.8&(1.85, 5.06)&1.5$\times 10^{-6}$\\ \hline
(5) (120, 250, 480, 380)&10&17.6&(0.97, 2.6)&3.8$\times 10^{-7}$\\ \hline
(6) (120, 250, 600, 450)&10&14.0&(1.03, 2.84)&4.6$\times 10^{-7}$ \\ \hline
(7) (120,250,700,550)&10&11.4&(0.72, 2.06)&2.6$\times 10^{-7}$\\ \hline
\end{tabular}\\

\noindent
Table 1. New contributions to $a_\mu$ and to $A(\tau\rightarrow\mu\gamma)$
due to $H_d$-$H_s$ loop in
ESSM. The masses $(M_1,M_2,M_E,M_{E'})$ are given in units of GeV. $E$ and
$E'$ are expected to be degenerate to within (10-50)\% at the electroweak
scale [see Eq. \eqref{eq:MEME}].\\

A glance at the table reveals the following features:

\noindent (1) For a decent range of the spectrum, with heavy leptons
$(E, E')$ having masses $\approx 300$-600 GeV (say) and thus the heavy quarks
having masses $\approx 700$-1500 GeV, and for reasonable positive values of
$\hat\lambda\approx 4$-10 [which would arise plausibly for
$\tan\beta\approx 3$-10 (say), see Eq. \eqref{eq:hatlambda}], ESSM provides
a sizeable positive contribution to
$a_\mu^{\rm ESSM}\approx (30$-$10)\times 10^{-10}$ (say).

\noindent (2) For the same range of the spectrum and the value of
$\hat\lambda$ as above, with
$a_\mu^{\rm ESSM}\approx (30$-$10)\times 10^{-10}$ (say), {\it ESSM makes
a correlated contribution to the branching ration for
$\tau\rightarrow\mu\gamma$}, which typically lies in the range of
(0.4-15$)\times 10^{-7}$. Given that the present experimental upper limit for
$B(\tau\rightarrow\mu\gamma)$ is around $10^{-6}$ \cite{PleData},
{\it ESSM quite reasonably predicts that $\tau\rightarrow\mu\gamma$ decay
should be discovered with a modest improvement of the current limit by a
 factor of 3-20}.  Studies at $B$--factories can be sensitive to the
level of few times  $10^{-8}$ in this branching ratios, while
LHC can probe even further.  

 It should be remarked that the ESSM-contribution to $a_\mu$ noted above is,
 of course, above and beyond the familiar SUSY-contribution to $a_\mu$ [10],
 which necessarily exists for ESSM as well. However, in the presence of
 $a_\mu^{\rm ESSM}$, even if the net new contribution to $a_\mu$ eventually
 needs to be in the range of (15-$30)\times 10^{-10}$ (say), bulk of this
 contribution can in principle come from $a_\mu^{\rm ESSM}$. That is if
 sleptons are not too light ($m_{\tilde l}\sim 400$ GeV, say) and if
 $\tan\beta$ is not too large ($\lesssim 10$, say), $a_\mu^{\rm SUSY}$ can
 be less than or of the order (5-10$)\times 10^{-10}$, while
 $a_\mu^{\rm ESSM}$ can be of the order (10-20$)\times 10^{-10}$ (say).
 [Depending upon the sign of the $\mu$ parameter, with $\hat\lambda >0$,
 the two contributions add or subtract.] However, in this case (i.e., with
 $a_\mu^{\rm ESSM}\sim (10-20)\times 10^{-10}$, $\tau\rightarrow\mu\gamma$
 should be discovered with the improvement in its current limit by a factor of
 3-20.

\subsection{Electric Dipole Moment of the Muon}

As noted in Sec. 5, the parameter $\hat{\lambda}$ in Eq. (16) is in
general complex, with its imaginary part being proportional to the
phases of the $A$--term and/or to the VEV $v_s$.  A complex $\hat{\lambda}$
will lead to a nonzero electric dipole moment (EDM) of the muon ($d_\mu$), 
arising from the same type of diagrams as in Fig. 1.  $d_\mu$ can be estimated 
to be $d_\mu \simeq a_\mu^{ESSM}/(2 m_\mu) {\rm arg}(\hat{\lambda})$.
This is in the range $(1-3) \times 10^{-22}{\rm arg}(\hat{\lambda})$ e-cm
for $a_\mu^{ESSM} = (10-30) \times 10^{-10}$.  The present experimental
limit on $d_\mu$ is $d_\mu \leq 10^{-18}$ e-cm.  There is a proposal
\cite{yannis} to improve this limit down to the level of $10^{-24}$ e-cm or
even $10^{-26}$ e-cm.  The ESSM framework presented here will predict
an observable signal in such experiments.  It should be emphasized
that the $H_d-H_s$ mixing contribution to the electron EDM ($d_e$)
is extremely small in our framework since the electron has highly
suppressed couplings to $(E,E')$ fields.  Thus the naive scaling
$d_e/d_\mu \sim m_e/m_\mu$ will not hold in our case (unlike in the
MSSM).  A linear scaling of the EDMs with the lepton mass would have
implied $d_\mu \leq 10^{-25}$ e-cm from the current limit on $d_e$
\cite{bb}.  

\subsection{${\bf b\rightarrow s\gamma}$}

Just like for $\tau\rightarrow\mu\gamma$, there would be new contributions
to the amplitude for $b\rightarrow s\gamma$ decay through $H_d$-$H_s$ loop
involving $D_{L,R}$ and $D'_{L,R}$ heavy quark exchanges (compare Fig. 1),
as well as through $W$ loop involving $(U,U')$-exchanges (compare Fig. 2).
We can simply obtain the new contributions to $b\rightarrow s\gamma$
amplitudes in ESSM by making the following substitutions in the
corresponding amplitudes for $\tau\rightarrow\mu\gamma$, listed in Eqs.
\eqref{eq:ALR}, \eqref{eq:ARL}, \eqref{eq:A1}, and \eqref{eq:A2}:
\begin{eqnarray}
\label{eq:subst}
A(\tau_{L,R}\rightarrow\mu_{R,L}\gamma)&\longrightarrow&
A(b_{L,R}\rightarrow s_{R,L}\gamma)\nonumber\\
(p_l,p'_l)&\longrightarrow&(p_d,p'_d)\nonumber\\
(M_E,M_{E'})&\longrightarrow&(M_D,M_{D'})\\
m_\tau &\longrightarrow& m_b \nonumber\\
Q_E^{\rm em}=Q_{E'}^{\rm em}=e&\longrightarrow&
Q_D^{\rm em}=Q_{D'}^{\rm em}=e/3 ~.\nonumber
\end{eqnarray}
As noted in Sec. \ref{sect2}, we have:
\begin{eqnarray}
\label{eq:pfactors}
\begin{array}{ll}
p_l=2(\eta+3\epsilon)\approx 0.27, & p_d=2(\eta-\epsilon)\approx -0.49\\
p'_l=2(\eta-3\epsilon)\approx -0.87, & p'_d=2(\eta+\epsilon)\approx -0.11
\end{array}
\end{eqnarray}
Using QCD renormalization factors for the effective momentum running from
the GUT to the electroweak scale, we get \cite{PS95}:
\begin{eqnarray}
M_{D,D'}=\eta_c M_{E,E'}
\end{eqnarray}
where $\eta_c\approx 2.8$. For an estimate, consider a relatively light
heavy lepton spectrum -- i.e., $M_{E,E'}\approx(420,$ 300) GeV with
$\hat\lambda=10$ [case (4) in Table 1], and thus
$M_{D,D'}\approx(1176,$ 840) GeV. Using the substitutions above, we get
$A_{R,L}^{H_dH_s}(b_{R,L}\rightarrow s_{L,R}\gamma)
\approx (8.5,35.8)\times 10^{-11}$ GeV$^{-1}\times(0.689)$, where the
factor (0.689) denotes the QCD renormalization of the effective operator
(see e.g., \cite{Buras}). Comparing with the Standard Model contribution
(see e.g., \cite{Buras,hewitt}) $A_R(b_R\rightarrow s_L\gamma)^{\rm SM}\approx
-\{(4G_F/\sqrt{2})(e/16\pi^2)V_{tb}V^*_{ts}\}\{2m_b\,c_7^{\rm eff}(\mu)\}
\approx -7.5\times 10^{-9}$ GeV$^{-1}$, where
$c_7^{\rm eff}(\mu)\approx -0.312$, we see that the new contributions due to
$H_d$-$H_s$ loop involving $(A_R^{H_dH_s},A_L^{H_dH_s})$ are only about
(0.6, 2.6)\% of the Standard Model contribution for $A_R$, which are thus too
small. As in the case of $\tau\rightarrow\mu\gamma$, the new contributions
involving $W$-loop are even smaller. The unimportance of the new
contributions to $b\rightarrow s\gamma$ amplitudes (in contrast to the
case of $\tau\rightarrow\mu\gamma$) arises primarily because of (i) the
difference in electric charges $Q_D/Q_E=1/3$, (ii) the heaviness of the
quark members $(D,D')$ compared to the leptonic members $(E,E')$ due to
QCD renormalization, and (iii) the difference in the $p$-factors
[see Eq. \eqref{eq:pfactors}].

\subsection{New Contributions to $\mu \rightarrow e\gamma$}

As explained in Sec. \ref{sect2}, if the entries in the upper 3$\times$3
block of Eq. \eqref{eq:Yuk} are set strictly to zero, one can always
go to the hat-basis in which the electron family would be completely
decoupled from the other four families ($\mu$, $\tau$, $Q$ and $Q'$) and
would remain massless. In this limit, the amplitude for $ \mu \rightarrow
e\gamma$ would of course vanish. The electron family does, however, get
masses and mixings with the other families owing to small entries $m_{ij}$
($\lesssim$ a few $MeV$) in the upper 3$\times$3 block of the mass-matrix,
which can arise through VEVs inserted into higher dimensional operators.
Given that there are new contributions to $a_\mu$ (i.e., to
$\mu_L \rightarrow \mu_R\gamma$) in ESSM especially from the $H_d$-$H_s$ loop,
which was evaluated in a basis where the muon is almost physical, except
for its small mixing with the electron, the amplitude for
$\mu \rightarrow e\gamma$-transition can be obtained simply by inserting
the $e$-$\mu$ mixing angles into $A(\mu_L\rightarrow \mu_R\gamma)$
\cite{FN11}. Thus we get [following the definition in Eq.
\eqref{eq:amplitude}]:
\begin{eqnarray}
\label{eq:amplitude5}
A(\mu_{L,R}\rightarrow e_{R,L}\gamma)\approx (e/2m_\mu)a_\mu^{\rm ESSM}
\Theta^{e\mu}_{R,L}
\end{eqnarray}
where $\Theta^{e\mu}_{R,L}\approx m^l_{12,21}/m_\mu$. Here $m^l_{12}$ and
$m^l_{21}$ are the $\bar{e}_R\mu_L$ and the $\bar{\mu}_Re_L$ mixing masses,
while $m^l_{11}$ (not shown) is the $\bar{e}_Re_L$ diagonal mass (all in the
hat basis), and $a_\mu^{\rm ESSM}$ is the contribution to $a_\mu$
from the $H_d$-$H_s$ loop, listed in Table 1 \cite{FN12}. [The amplitude
$A(\mu\rightarrow e\gamma)$ would also get contributions by inserting
$e$-$\tau$ mixing angles -- i.e., $\Theta^{e\tau}_{R,L}
 \approx m^l_{13,31}/m_\tau$ -- into $A(\tau\rightarrow \mu\gamma)$
 [given by Eqs \eqref{eq:ALR} and  \eqref{eq:ARL}]. One can
 estimate (using obvious notation) that $A(\mu_L\rightarrow e_R)_
 {\tau \rightarrow \mu}/ A(\mu_L\rightarrow e_R)_
{\mbox\scriptsize
\mbox{Eq. \eqref{eq:amplitude5}}}\sim (2/p)(1/2.5)(m_\mu/m_\tau)
 (m_{13}/m_{12})\approx (1/5.6)(m_{13}/m_{12})$. The analogous ratio for
$A(\mu_R\rightarrow e_L)$ is $\approx -(1/30) (m_{31}/m_{21})$. Thus, barring
accidental cancellation in both channels, which is unlikely, it should
suffice as an estimate to include only the contribution shown in Eq.
\eqref{eq:amplitude5}, which should yield the right magnitude within a factor
2-3 (say).] Using Table 1 as a guide and setting $a_\mu^{ESSM}\equiv x_\mu
(30\times 10^{-10})$, and furthermore assuming for simplicity
$\Theta_L^{e\mu}\approx \Theta_R^{e\mu}\equiv \Theta^{e\mu}$, we get [using
Eq. \eqref{eq:amplitude5}]:
\begin{eqnarray}
\label{eq:Gam}
\Gamma(\mu \rightarrow e\gamma)^{Th}=K(8\times 10^{-22}\; GeV)
(x_\mu \Theta^{e\mu})^2~.
\end{eqnarray}
Here $K$ denotes a correction factor of order one ($K\approx$ 1/4 to 4, say),
which can arise by allowing for contribution from $\tau \rightarrow \mu
\gamma$ transition and for $\Theta_L^{e\mu}\neq\Theta_R^{e\mu}$, etc. The
experimental limit $B(\mu\rightarrow e\gamma)<1.6\times 10^{-11}$
\cite{PleData} thus provides an upper limit on $m^l_{12}\sim m^l_{21}$
(with $K \geqslant$ 1/4, say) given by:
\begin{eqnarray}
\label{eq:Memu}
m_{e\mu}\lesssim (1/65,\; 1/42,\; 1/32,\; 1/22,\; 1/11)\; MeV
\end{eqnarray}
for $a_\mu^{ESSM}=$(30, 20, 15, 10, 5)$\times 10^{-10}$. Here $m_{e\mu}$ may
be viewed roughly as the average of $m^l_{12}$ and $m^l_{21}$.

Apriori, one might have expected $m_{e\mu}$ to be at least of order
$m_{e}\approx 1/2\; MeV$ (if not of order $m_\mu\sqrt{m_e/m_\mu}$) which
is, however, a factor (30 to 10) higher than the values shown in
Eq. \eqref{eq:Memu}. In this sense, if a sizeable contribution to $a_\mu$
($\gtrsim 15\times 10^{-10}$, say) should come from the $H_d$-$H_s$ and
$W$-loops in ESSM, a natural explanation for the large suppression of
$m_{e\mu}$, as required by the limit $\Gamma(\mu \rightarrow e\gamma)$,
would clearly be warranted. While this is a burden on ESSM, we should remark
that given the smallness of the elements of the mass-matrix involving the
first family, it is difficult to pin-down the SO(10)-structure of the
corresponding Yukawa couplings, because these may arise from a variety of
 SO(10)-invariant higher dimensional operators. In fact, consistent with
SO(10)-invariance, there can exist a mechanism \cite{FN13} (analogous to that
of the doublet-triplet splitting in SO(10) \cite{MechanismSplit}), which
would contribute to mixings of the first family with the other two, only
in the quark-sector, but not in the lepton-sector. This could retain the
successes of Ref. \cite{BPW} as regards the predictions of the Cabibbo
angle, $V_{ub}$ and $m_d$, while yielding vanishing $e$-$\mu$ and $e$-$\tau$
mixings, and simultaneously $m_e \neq 0$. Details of this discussion will be
presented in a separate paper.

In spite of this specific mechanism, however, it is hard to see why
$m_{e\mu}$ and thus $\Theta_{e\mu}$ should be strictly zero. Even if
$m_{e\mu}$ is suppressed by a factor of 50 to 100, say, compared to $m_e$
(and that seems to be rather extreme), with
$a_\mu^{ESSM}\gtrsim 10\times 10^{-10}$ and $K \gtrsim 1/4$
[see Eq. \eqref{eq:Gam}], we would expect:
\begin{eqnarray}
\label{eq:B}
B(\mu\rightarrow e\gamma)\gtrsim 1.6\times 10^{-11}(1/14\mbox{ to }1/56)~.
\end{eqnarray}
In short, within ESSM, the decay $\mu\rightarrow e\gamma$ is generically
expected to occur at a decent level so that it should have been seen already.
Even with a rather pessimistic scenario for $m_{e\mu}$ as mentioned above,
the decay should be seen with an improvement in the current limit by a
factor of 5 to 50 (say), especially if $a_\mu^{ESSM}\gtrsim 10\times
10^{-10}$.

\section{Concluding Remarks}

The ESSM framework we have adopted here has been motivated on several
grounds, as noted in our earlier papers \cite{JCPBabu,BabuJCP} and 
summarized here in Sec. 2. ESSM has been embedded into an SO(10)
unified theory which makes correlations among several observable
quantities (such as those between $\tau \rightarrow \mu \gamma$,
$b \rightarrow s \gamma$ and neutrino oscillations) possible.  Such
an embedding preserves the unification of gauge couplings and
provides a quantitative understanding of the pattern of quark and
lepton masses, including the smallness of $V_{cb}$ and the largeness
of the $\nu_\mu-\nu_\tau$ oscillation angle.

In this paper, we have studied the new contributions of ESSM to
radiative processes including $\tau \rightarrow \mu \gamma$, $b \rightarrow
s \gamma$, $\mu \rightarrow e \gamma$, $(g-2)_\mu$ and the muon EDM.
We have shown that ESSM makes significant contributions especially to
the decays $\tau \rightarrow \mu \gamma$ and $\mu \rightarrow e \gamma$
and simultaneously to $(g-2)_\mu$.  For a large and plausible range of
the relevant parameters (see Table 1), we obtain $a_\mu^{ESSM} \approx
+(10-30) \times 10^{-10}$, and predict that $\tau \rightarrow \mu \gamma$
should be discovered with an improvement in the current limit by 
a factor of 3-20.  The implication for the discovery of $\mu \rightarrow
e \gamma$ is very similar.  The EDM of the muon is expected to be in
the range of $10^{-22}$ e-cm, which should be accessible to the next
generation of experiments.  Thus radiative processes can provide an
effective probe of ESSM before a direct search for the heavy fermions
is feasible at the LHC.  The hallmark of ESSM is of course
the existence of complete
vectorlike families ($U,~D,~N,~E)_{L,R}$ and ($U',~D',~E',~N')_{L,R}$
with masses in the range of 200 GeV to 2 TeV )say), which will certainly
be tested at the LHC and a future linear collider.

\section*{Acknowledgements}

JCP wishes to thank Susan Gardner, Gudrun Hiller and Kirill Melnikov for
helpful discussions.  He also wishes to acknowledge the hospitality
of the SLAC theory group during his sabbatical visit there.  The 
authors wish to acknowledge  the hospitality of the theory group
at CERN where this work was completed.  The research of KSB is supported in 
part by the US Department
of Energy Grant No. DE-FG03-98ER-41076, DE-FG03-01ER45684 and by
a grant from the Research Corporation.  That of JCP is supported in
part by DOE Grant No. DE-FG02-96ER-41015.

}

\newpage

\begin{figure}
\begin{center}
\mbox{\epsfig{file=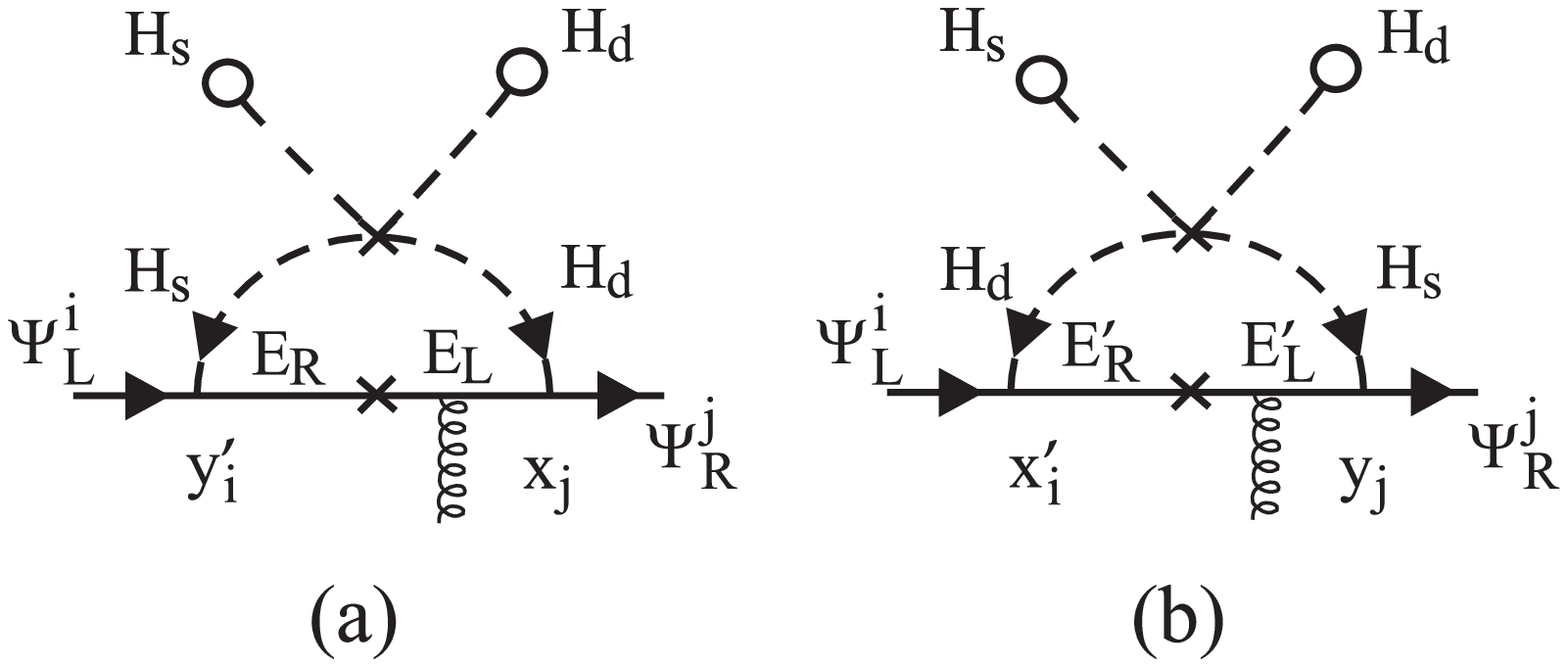,width=160mm}}
\caption{Contributions to $\ell_i \rightarrow
\ell_j \gamma$ arising through $H_d$-$H_s$ mixing}
\end{center}
\end{figure}

\begin{figure}
\begin{center}
\mbox{\epsfig{file=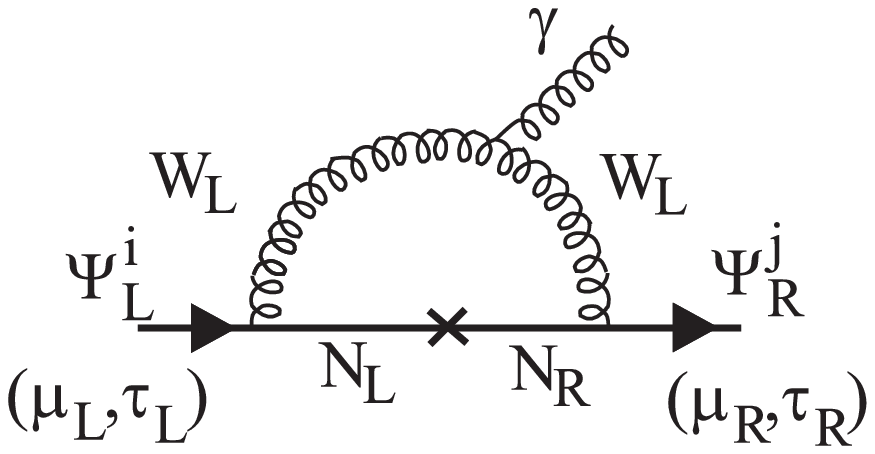,width=80mm}}
\caption{New contributions to radiative transitions from $W$-loop}
\end{center}
\end{figure}


\begin{thebibliography}{23}
\bibitem{FN11new} To obtain $a_\mu^{\rm SUSY}\approx 25\times 10^{-10}$ (say),
one would, however, need  rather light sleptons
($m_{\tilde l}\approx 200$-400 GeV) and $\tan\beta$ to be appropriately
large ($\approx 6$-24). Such light sleptons and correspondingly light squarks
($m_{\tilde q}\approx 400$-800 GeV, say), together with moderately large
$\tan\beta$, could, however, pose theoretical dilemmas pertaining to issues
of (a) proton decay within SUSY grand unification, and possibly also (b)
edm's of the electron and the neutron.
\bibitem{Barr90} This mechanism for inter-family mass hierarchy was also
noted by S. Barr, Phys. Rev. {\bf D42}, 3150 (1990).  In this work the
vector-like families were assumed to have GUT scale masses, which would
not be relevant for present discussions.
\bibitem{NewFN} Masses and mixings of the first family fermions
(in the hat basis) are attributed to higher dimensional operators
consistent with flavor symmetries which directly provide entries
of order 1 MeV in the $3 \times 3$ sector involving the chiral
families.
\bibitem{FN2} The case of MSSM embedded in SO(10), considered in
Ref. \cite{BPW}, leads to a value for $m_b$, which is about 10 \%
higher than the observed value. Thus ESSM (with $\kappa=\delta=0$) would lead 
to even larger values for $m_b$.
\bibitem{FN3} With $\kappa\sim 1/10$, $M_E$ can be lower
than $M_{E'}$ by (say) 10-20 \% at the electroweak scale, but the sign
of the departure from unity in Eq. \eqref{eq:MEME} will not alter our
discussion of $\tau\rightarrow\mu\gamma$ (see next section).
\bibitem{FN4} In addition, with $\kappa=\delta=0$, there would be equalities
between quark and lepton couplings, so that $z_u=z_d=z_l=z_\nu$;
$x_u=x_l=x_d=x_\nu$, etc. and $M_U=M_D=M_{U'}=M{D'}=M_E=M_{E'}$ at the
GUT-scale.
\bibitem{FN5} Since $\tau\rightarrow\mu\gamma$ is rather sensitive to
degree of non-degeneracy of $E$ and $E'$, we would still permit
$M_E$ and $M_{E'}$ to receive small contributions from the $\kappa$-term as
suggested by $m_b/m_\tau$. So, together with renormalization effects
$(M_E/M_{E'})$ may vary in a range, such as that given by
Eq. \eqref{eq:MEME}. Allowing for the $\kappa$-term would, however, not have 
a major effect on the relevant mixing angles.
\bibitem{FN6} This term is analogous to or could play the role of that in
the Next to Minimal Supersymmetric Standard Model (NMSSM),
which gives rise to the $\mu$-term.
Using $\langle H_s \rangle=v_s$, one would get $\mu=\lambda v_s$.
\bibitem{FN34} This is because, (a) the $Z^0$-current has a partial GIM-like
invariant form in the subspaces of $(\mu_L,\tau_L,E_L,E_R)$ and
 $(\mu_R,\tau_R,E'_L,E'_R)$, and thus the new contributions involving
 physical $(E,E')$ in the loop and physical $(\mu,\tau)$ as external
 particles vanish except for $E-E'$ mixing which is small
 $(\simeq x_3y'_3v_dv_s/v_0^2\lesssim (v_d/v_0)(v_s/v_0)<1\%)$; and in
 addition (b) the effective coupling of $Z^0$ given by $(g_2/\cos \theta_W)
 (I_{3L}-Q_{\rm em} \sin^2\theta_W)$ is smaller than that of $W^+$.
\bibitem{FN11} One would of course get essentially the same answer if one
calculates in the true physical basis for all fields, because the latter
would induce non-diagonal couplings of the electron to the physical
($E$, $E'$) and the N-fields, with the emission of ($H_d$-$H_s$) and
$W$-fields respectively.
\bibitem{FN12} $a_\mu^{\rm ESSM}$ in Eq. \eqref{eq:amplitude5} does not, of
course, include other contributions to $a_\mu$ such as those from
supersymmetry which do not contribute to off-diagonal transitions.
\end{thebibliography}
\end{document}